 \documentclass[superscriptaddress,nofootinbib]{revtex4}
\usepackage{amsfonts}
\usepackage{amssymb}
\usepackage{amsmath}
\usepackage{amsthm}
\usepackage{epsfig}
\usepackage{marginnote}
\newcommand{\be}{\begin{eqnarray}}
\newcommand{\ee}{\end{eqnarray}}
\newcommand{\la}{\langle}
\newcommand{\ra}{\rangle}

\begin{document}
\title{\bf Mass spectrum of pseudo-scalar
glueballs from a  Bethe-Salpeter  approach
 with the rainbow-ladder truncation}
 \author{L.~P. Kaptari}
\affiliation{Bogoliubov Lab.~Theor.~Phys., 141980, JINR, Dubna, Russia }
\affiliation{Helmholtz-Zentrum Dresden-Rossendorf, PF 510119, 01314
Dresden, Germany}
\author { B.~K\"ampfer}
\affiliation{Helmholtz-Zentrum Dresden-Rossendorf, PF 510119, 01314
Dresden, Germany}
\affiliation{Institut f\"ur Theoretische Physik, TU Dresden, 01062 Dresden, Germany}

 \begin{abstract}
We suggest a framework based on the rainbow approximation to the Dyson-Schwinger and Bethe-Salpeter
equations with effective parameters  adjusted to lattice QCD data to calculate the masses of the ground and excited states of pseudo-scalar glueballs. The structure of the truncated Bethe-Salpeter equation with the gluon and ghost propagators as solutions of the truncated Dyson-Schwinger equations is analysed in Landau gauge. Both, the Bete-Salpeter and Dyson-Schwinger equations, are solved numerically within the same rainbow-ladder truncation with the same effective parameters which ensure consistency of the approach. We found that with a set of parameters, which provides a good description of the lattice data within the Dyson-Schwinger approach, the solutions of the Bethe Salpeter equation for the pseudo-scalar glueballs exhibit a rich mass spectrum which also includes  the ground   and excited states predicted by lattice calculations. The obtained mass spectrum contains also several intermediate excitations beyond the lattice approaches. The partial Bethe-Salpeter amplitudes of the pseudo-scalar  glueballs  are presented as well.
  \end{abstract}
%

 \maketitle

\section{Introduction}
The Quantum Chromodynamics (QCD), the fundamental theory of strong interaction, essentially relies on
the exact SU(3)-color symmetry, according to which the gluons, as gauge fields, carry color charges and
are allowed to interact among themselves. Consequently,  gluons can form
pure gluonic bound states, also referred to as glueballs. The occurrence of
 glueballs is one of the early predictions of the  strong interaction phenomena described by
  QCD~\cite{mink,Jaffe}. Experimental discovery of glueballs would be a formidable confirmation of the validity of
theoretical approaches to the nonperturbative QCD. Although there is an intense experimental effort
to detect glueballs, for the moment there is no direct and  unambiguous evidence of them,
cf. Ref.~\cite{ulrich,Jia:2016cgl}. Possible reasons for this is that it is not possible to distinguish the glueballs $(gg)$
  from conventional $(q\bar q)$ mesons only by quantum numbers and masses.  There are needs for other
 more specific tools to investigate glueballs, such as investigation of
meson mixing,  flavor independent decay processes, life-time etc.
 Therefore, the  study of glueballs is among the most interesting and
challenging problems intensively studied by theorists and experimentalists;
 a bulk of the running and projected   experiments   of the  research centers, e.g.
 Belle (Japan),  BESIII (Beijing, China),
 LHC (CERN), GlueX (JLAB,USA), NICA (Dubna, Russia), HIAF (China),  PANDA at FAIR/GSI (Germany) etc.,
 include  in the research programs
comprehensive investigations of possible manifestations of glueballs.
Theoretical frameworks such as the flux tube model~\cite{Robson:1978iu,Isgur:1984bm},
constituent models~\cite{Jaffe,Carlson:1984wq,Chanowitz:1982qj,Cornwall:1982zn,Cho:2015rsa,Boulanger:2008aj},
holographic approaches~\cite{viena,Bellantuono:2015fia,Chen:2015zhh,Brunner:2016ygk}, and
approaches based on QCD Sum Rules~\cite{Shifman:1978bx,Shuryak:1982dp,kolya,KolyaPRL,kolya1} have
shed some light on the potential identification of experimental states dominated or
partially governed by glueball components. Also, numerous simulations of lattice QCD seems to confirm
the existence of ground and exited glueball states with masses
below 5~GeV~\cite{Albanese,Chen,Morningstar,Gabadadze}
(for a more detailed  review see Ref.~\cite{Glueballstatus} and references therein).

 Another interesting problem is the glueball-meson
mixing in the lowest-lying scalar mesons.  The question whether the lowest-lying scalar mesons
are of a pure quarkonium nature, or  there are   mixing  phenomena of glueball  states~\cite{mixing}
remains still open.
To solve these problems one needs to develop models  within which it becomes possible to investigate, on a
common footing, the glueball masses, glueball wave functions, decay modes and constants,  etc.
Such approaches can be based on the combined Dyson-Schwinger (DS) and Bethe-Salpeter (BS)
formalisms, cf. Refs.~\cite{glueBS,GlueBallBSPRD87}.

In the present paper we suggest an approach, similar to the rainbow  Dyson-Schwinger-Bethe-Salpeter
model for quark-antiquark bound
states~\cite{rob-1,ourFB,Maris:2003vk,Alkofer,fisher,rob-2,dorkinBSmesons,OurAnalytical},
to solve the truncated Bethe-Salpeter equation (tBSE) for
two-gluon systems. According to the classification of
 two-photon (two-gluon, colorless) bound states~\cite{landau}, the simplest, and at the same time, the lightest
 glueballs are the scalar ($0^{++}$) and pseudo-scalar ($0^{-+}$) states.
 We focus our attention on the pseudo-scalar glueballs. From theoretical point of view,
 the pseudo-scalar glueballs are less complicate. However, even in this case
 the theoretical treatment turns out to be rather cumbersome and involved.

 The key property of the presented framework is the self-consistency of the  treatment of the
 quark and gluon propagators in both, truncated Dyson-Schwinger (tDS)  and
 truncated Bethe-Salpeter (tBS) equations by employing in both cases the same
 effective interaction kernel.

  Since  the  momentum dependence of the gluon and ghost dressing functions,  the tBS equation requires
  an analytical continuation of the gluon and ghost propagators in the
  complex plane of Euclidean momenta which can be achieved either by corresponding numerical
  continuations of the solution obtained along the positive real axis or by solving directly
  the tDS equation in the complex domain of validity
  of the equation itself.  For this one needs first to solve the tDS equation along the real axis, then
  by using the same effective parameters, to find the gluon propagators in complex Euclidean space.
 In~\cite{EJPPlus} we  analysed preliminarily the prerequisites to solve the tDS equation  along the real axis and
  investigated the analytical properties of the complex solution for the
  gluon and ghost propagators in complex Euclidean space. The present
  paper is a continuation of the previous studies~\cite{ourFB,dorkinBSmesons,OurAnalytical,EJPPlus}
    of the tDS and tBS equations, now with the scope of studying the pseudo-scalar glueballs
    within the rainbow-ladder truncation with the gluon propagators  previously obtained  in
    Ref.~\cite{EJPPlus}. Note that in~\cite{EJPPlus} the effective parameters for the
    tDS equation have been adjusted to obtain a reasonable agreement with lattice SU(2) calculations for the
    gluon and ghost propagators without any connection to the possible gluon bound states.
     In the present paper we re-analyse the effective rainbow parameters in
    order to achieve simultaneously  a better description of the lattice  data for propagators
    and to obtain  a realistic description of the mass spectrum of the pseudo-scalar glueballs.

 Our paper is organized as follows. In Sec.~\ref{s:bse}, Subsecs.~\ref{Bet}~and~\ref{Dys},
 we briefly discuss the  tBS and tDS equations, relevant to describe a glueball as two-gluon
 bound states. The numerical solutions of the tDS equations with the re-fitted parameters
  together with comparison with lattice QCD data are presented in Subsecs.~\ref{Dys} and \ref{dysBow}.
 The explicit expressions for the BS amplitude within the rainbow approximation are presented
 in Sec.~\ref{bow}. Details of numerical calculations are presented in  Section~\ref{num}:
 in Subsec.~\ref{sec:complex} we discuss the procedure of finding the complex solution for tDS equations and,
 in Subsec.~\ref{numBS} we briefly discuss the numerical algorithm used to  solve the tBS equation.
 Conclusions and summary are collected in Sec.~\ref{summary}. In Appendix~\ref{A} and \ref{B}
 details of analytical computation of the relevant angular integrations are discussed.

\section{Bethe-Salpeter Equation for glueballs}

\label{s:bse}
The combined Dyson-Schwinger--Bethe-Salpeter approach used in the present paper
  to  describe a glueball as bound state of two dressed gluons, implies the self-consistent
   treatment of the gluon propagator in both, tBS  and tDS, equations. It means that all
   ingredients for the corresponding diagrams ($3g$-vertex functions, effective form factors, gluon propagators,
   normalization scale etc) are the same. In the following we work along this strategy,
i.e. we  elaborate an effective model  within which  (i) the solution of the gluon and ghost propagators,
consistent with lattice  data, is obtained on the positive real axis
 of the momentum, (ii) then  the real solution is generalized
 for  complex momenta, relevant to the domain in
 Euclidean space where the tBS equation is defined, and
 (iii) solve the tBS equation to obtain the partial Bethe-Salpeter amplitudes for the glueball.

\begin{figure}[!ht]
  \begin{center}
 \includegraphics[scale=0.4 ,angle=0]{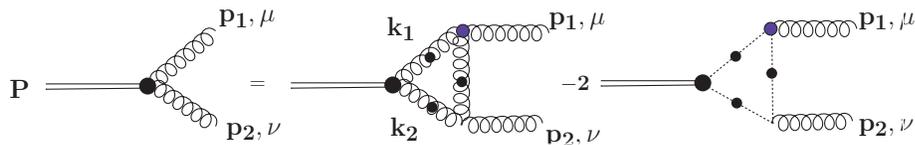}\vspace*{3mm}
  \end{center}
 \caption{ Diagrammatic representation  of the Bethe-Salpeter  equation for gluon (wiggly lines)
   bound states (double line).  The irreducible one-particle vertices and the full propagators are represented and marked by filled blobs.}
 \label{Bseq}
 \end{figure}

  In the present paper we focus our attention on the simplest gluon bound states, namely on
  pseudo-scalar pure glueballs. In this case only the first r.h.s. diagram in Fig.~\ref{Bseq} contributes
  to  the amplitude. In case of  scalar glueballs, besides the two terms r.h.s. in   Fig.~\ref{Bseq}, also diagrams
 which couple the ghost amplitude with the glueball ones, must
  be taken into account. This makes  investigations  of scalar glueballs
  much more complicated (see e.g. Ref.~\cite{GlueBallBSPRD87} for some details)
  and cumbersome for numerical calculations.

 \subsection{BS amplitude for Pseudoscalar glueballs} \label{Bet}

The BS amplitude of a colorless pure glueball with total spin and parity $J^\pi$ and total momentum
$P$ is defined in the standard way,
 \be
 A^{\mu\nu}(x_1,x_2) =  \left\la 0 \left | T\left[\hat A^\mu(x_1)\hat A^\nu(x_2)\right]
 \right |J^\pi,P\right\ra ,\label{BSA}
 \ee
 where, for brevity, the color indices of the gluon field operators $\hat A_{\mu,\nu}(x)$ are suppressed.
  Usually one considers the Fourier transform $A^{\mu\nu}(p_1,p_2)$ of the amplitude (\ref{BSA}), which due to    translation invariance, depends on the relative momentum $p=(p_1-p_2)/2$ and the total momentum, $P=(p_1+p_2)$,
   of the glueball. By definition, the amplitude  $A^{\mu\nu}(p_1,p_2)$ is  transverse
\be
p_{1\mu}A^{\mu\nu}(p_1,p_2)=A^{\mu\nu}(p_1,p_2)p_{2\nu}=0.
\label{gauge}
\ee
Often, instead of  the BS amplitude $A^{\mu\nu}(p_1,p_2)$ one considers  the BS
vertex function $G_{\alpha\beta}(p_1,p_2)$ defined as
 \be
 A^{\mu\nu}(p_1,p_2)=D^{\mu\alpha}(p_1)G_{\alpha\beta}(p_1,p_2)D^{\beta\nu}(p_2). \label{vertex}
\ee
From (\ref{gauge}) it follows that the BS vertex  $G_{\alpha\beta}(p_1,p_2)$  is also transverse
\be &&
p_{1\mu}A^{\mu\nu}(p_1,p_2)=p_{1\mu} D^{\mu\alpha}(p_1)G_{\alpha\beta}(p_1,p_2)D^{\beta\nu}(p_2)=
-i\xi \frac{p_{1\alpha}}{p_1^2}
G_{\alpha\beta}(p_1,p_2)D^{\beta\nu}(p_2)=0,\nonumber \\ &&
A^{\mu\nu}(p_1,p_2)p_{2\nu} = -i\xi D^{\mu\alpha}(p_1)  G_{\alpha\beta}(p_1,p_2)
\frac{p_{2\beta}}{p_2^2} =0, \label{gauge1}
\ee
where $\xi$ is the gauge parameter of the gluon propagator,
$D^{\mu\nu}(p)=-i \frac{Z(p)}{p^2}
\left( g^{\mu\nu}-\frac{p^\mu p^\nu}{p^2}\right )
-i\xi\frac{p^\mu p^\nu}{p^4}$, where $Z(p)$ is the corresponding dressing functions.
In what follows we work in Landau gauge, $\xi\to 0$. From (\ref{gauge1})  a useful relation follows:
\be
A^{\mu\nu}(p_1,p_2)=D^{\mu\alpha}(p_1)G_{\alpha\beta}(p_1,p_2)D^{\beta\nu}(p_2)=
-\frac{Z(p_1^2)Z(p_2^2)}{p_1^2 p_2^2}  G^{\mu\nu}(p_1,p_2).
\ee

With these preliminary notations, the BS amplitude and BS vertex
for a pseudo-scalar glueball (the first r.h.s. diagram  in Fig.~\ref{Bseq}) read as
\be
A_{\mu\nu}(p_1,p_2) =   D_{\mu\mu_1}(p_1)  \left( -N_c\int \frac{d^4 k}{(2\pi)^4 }
 \Gamma_1^{\mu_1 \alpha\lambda}(p_1,k_1,\kappa) A_{\alpha\beta}(k_1,k_2)
\Gamma_2^{\nu_1 \beta \lambda_1}(p_2,k_2,\kappa) D_{\lambda\lambda_1}(\kappa) \right)
D_{\nu\nu_1}(p_2),  \nonumber \\&& \label{amplBS}
\ee
for the amplitude, and
\be &&
G_{\mu\nu}(p_1,p_2)=-\left[ \frac{Z_1(p_1)Z_2(p_2)}{p_1^2p_2^2}\right]^{-1} A_{\mu\nu}(p_1,p_2)=\nonumber\\ && =
t_{\mu\mu_1}(p_1)\left(-N_c\int\frac{d^4 k}{(2\pi)^4} \frac{Z_1(k_1^2)Z_2(k_2^2)}{k_1^2k_2^2}
\Gamma_1^{\mu_1\alpha\lambda}(p_1,k_1,\kappa)G_{\alpha\beta}(k_1,k_2) \Gamma_2^{\nu_1,\beta\lambda_1}
(p_2,k_2,\kappa) D_{\lambda\lambda_1}(\kappa)\right ) t_{\nu\nu_1}(p_2),
\nonumber \\ && \label{gammaBS}
\ee
for the BS vertex. In the above equations, $\Gamma_1^{\mu_1\alpha\lambda}(p_1,k_1,\kappa)$
and $\Gamma_2^{\nu_1,\beta\lambda_1}
(p_2,k_2,\kappa)$ are the $3g$-vertices corresponding to the  first r.h.s. diagram in Fig.~\ref{Bseq},
 and $t_{\mu\nu}(p)\equiv\left( g_{\mu\nu}-\displaystyle\frac{p_\mu p_\nu}{p^2}\right)$
 denotes the transverse projection operator.
Observe that, due to the presence of $t_{\mu\nu}(p_{1,2})$  both,
the   BS amplitude (\ref{amplBS}) and the BS vertex function (\ref{gammaBS}),
are manifestly transverse. In principle, these two equations are completely equivalent.
The only difference is that
in the equation for the BS amplitude the two gluon propagators are
outside the loop integral, while for the BS vertex functions the gluon propagators
are subjects of a four-dimensional integration. As mentioned above  consistency of the approach requires that
the gluon propagators in (\ref{amplBS})-(\ref{gammaBS})  are  solutions of the  tDS equation  obtained
within the same approach as the tBS equation.
As a rule, even in the simplest case, the tDS equation is solved numerically.
This causes additional difficulties in (\ref{amplBS})-(\ref{gammaBS})
when carrying out the  angular integrations. However, since in eq.~(\ref{amplBS})
the numerical solutions of the tDS equation are  outside the integral,
such numerical problems can be essentially  minimized. Moreover,  employing a specific form of the
phenomenological interaction kernel in (\ref{amplBS}) all angular integrations, over the spacial
and hyper angles of the integration momentum $k$,  can be performed
 analytically  (see below). This is the reason to  consider  the BS amplitude rather than the
  vertex (\ref{gammaBS}).

Prior to proceed with calculations of the amplitude (\ref{amplBS}) we come back to the
solutions  of the tDS equations, reported in Ref.~\cite{EJPPlus},
and re-analyse the effective parameters of the model in the context of a simultaneous description
  of the gluon propagators from  tDS  and the BS amplitude from tBS equations.
\subsection{Coupled Dyson-Schwinger equations for gluons and ghosts} \label{Dys}
In most approaches, the tDS equation usually is solved numerically by implementing different approximation
schemes. The simplest one
consists in a  replacement of the fully dressed three-gluon and  ghost-gluon vertices by their bare values,
a procedure known as the Mandelstam approximation~\cite{CPCMandelstam,MANDELSTAM_Approx,PennigtonMandels}
and the y-max approximation~\cite{YmaxApprox}. In order to simplify the
angular integration, in the Mandelstam approximation the gluon-ghost coupling is  neglected.
 In Ref.~\cite{YmaxApprox} the coupling of the gluon to the ghost was considered,
 however  additional simplifications for the gluon, $Z(k^2)$, and ghost, $G(k^2)$, dressing functions
  have been introduced, again  to  facilitate the angular integrations  and the analytical and numerical
analysis of the resulting equations. A more rigorous analysis of the tDS equation has been presented  in a series of
 publications (see, e.g. Refs.~\cite{IRGluonProp_AlkoferSmekal,FischerPhD,SmekalAnnPhys,Fischer_Alkofer_Reinhard}
 and references therein), where much attention has been focused on a detailed investigation
 of the gluon-gluon and ghost-gluon vertices and on the  implementation of
 the Slavnov-Taylor identities for these  vertices.
 With some additional approximations the infrared behavior of gluon and ghost propagators
has been obtained analytically and compared with the available lattice QCD calculations.
In Ref.~\cite{PawlowskyFicher} a thorough analysis of the relevance of the Slavnov-Taylor identities,
 renormalization procedures and divergences in the tDS equation is presented in some detail.
 Comparison of the numerical calculations
for the gluon and ghost dressing functions and  running coupling $\alpha_s$ with lattice data have
been presented as well. Similar calculations together with a comparison with  lattice data are
presented also in Ref.~\cite{IRGreen_FewBody2012}
(for a more detailed review see Ref.~\cite{FisherReview} and references therein quoted).

In~\cite{EJPPlus} we suggested an approach based on the rainbow approximation
to solve the tDS   system of equations for gluon and ghost dressing functions. It has been shown that it
is possible to establish a set  of effective parameters to describe reasonable well the lattice
SU(2) data.  Also, it has been mentioned that such a set of parameters is not unique;
one can find several different sets of  parameters
which  also  provide good descriptions of data. Recall that, in case of quarkonia (mesons)
the effective rainbow
parameters have been fitted  also to describe the lowest quark-antiquark bound states
(pions) and the quark-antiquark condensate. Contrary to this case,
in Ref.~\cite{EJPPlus}  the parameters have been adjusted
to lattice data solely for the propagator functions without any connection to possible bound states.
Here we come back to the tDS equation and  re-fit the parameters  with the scope of providing simultaneously
a good solution for the gluuon and ghost propagators and   reasonable results for
  the ground state of the pseudo-scalar glueballs.

 \begin{figure}[!ht]
 \begin{center}
 \includegraphics[scale=0.55 ,angle=0]{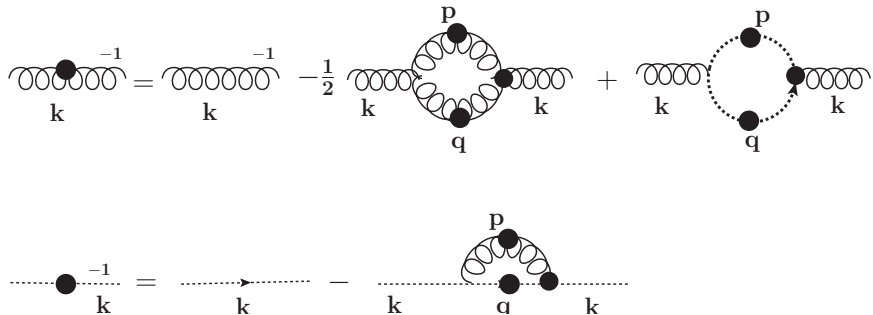}\vspace*{3mm}
 \end{center}
 \caption{ Diagrammatic representation  of the employed tDS   equations for gluon (top line) and
  ghost (bottom line) propagators. The internal wiggly and  dashed  lines denote
   the full propagators marked by filled blobs. The irreducible one-particle
   vertices are also denoted by filled blobs. In the gluon tDS  equation,
    terms with four-gluon vertices and quark loops  have been dismissed.}
 \label{dse}
 \end{figure}
\subsection{tDS equation within the rainbow approximation}\label{dysBow}

\begin{figure}[!ht]
 \begin{center}
 \includegraphics[scale=0.55 ,angle=0]{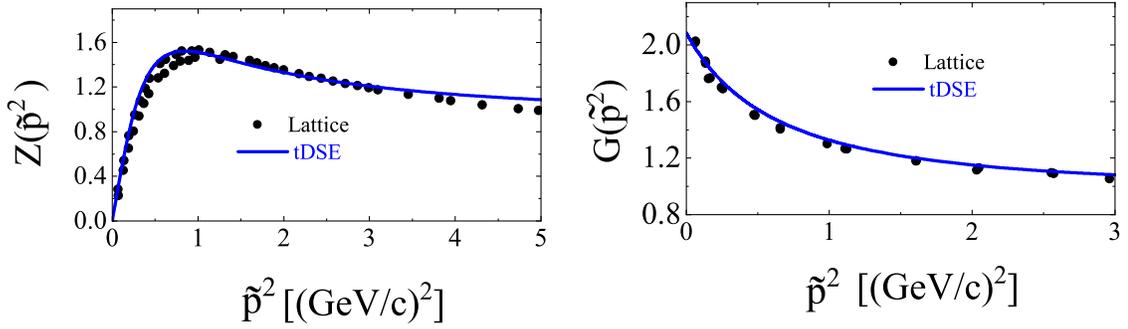}
 \end{center}
 \caption{(Color online) Solution of the  tDS equations  (solid lines) in comparison
with lattice SU(2) calculations~\cite{BornyakovLattice,GhostLatticeMishaPRD}  (filled circles). Left panel:
gluon dressing function, right panel: ghost dressing function.}
 \label{dsress}
 \end{figure}

\begin{figure}[!ht]
 \begin{center}
 \includegraphics[scale=0.4 ,angle=0]{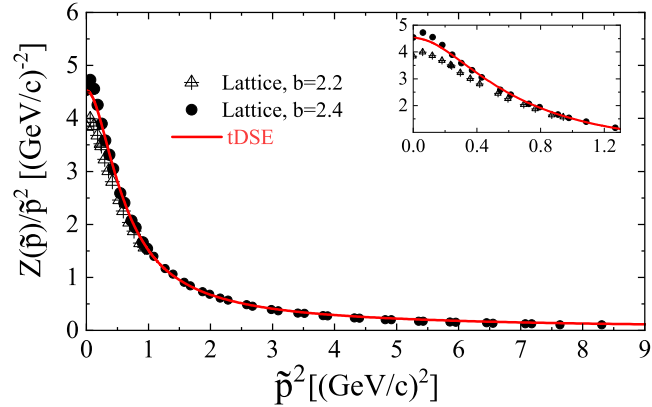}
 \end{center}
 \caption{(Color online) Solution of the  tDS equations  (solid lines) for the
gluon propagator in comparison
with lattice SU(2) data~\cite{BornyakovLattice,GhostLatticeMishaPRD}  (filled circles and symbols).
The inset is a zoom into the infra-red region. }
 \label{props}
 \end{figure}

  Diagrammatically, the  system of coupled tDS equations for the gluon and
 ghost propagators is presented  in Fig.~\ref{dse}. The explicit expressions
 for these diagrams have been computed within the rainbow approximation and presented in some details
 in Ref.~\cite{EJPPlus}. Since we are interested in bound states, where the main contribution
 comes from the infra-red region, particular attention in Ref.~\cite{EJPPlus} has been paid to
 the conjecture about the behaviour of the gluon dressing function $Z(p^2)$ and ghost
 dressing function $G(p^2)$ at the origin. So, if one adopts finiteness  of the ghost dressing  and $Z(p^2)\sim p^2$, then  a
 family of the so-called decoupling  solutions are generated, cf.~\cite{PawlowskyFicher}. Contrarily,
 the scaling solutions  imply a divergent ghost dressing ($G(0)\to\infty$) and
 $Z(p^2)\sim (p^2)^{\kappa} $ with $0.5< \kappa <1 $ at the origin~\cite{SmekalAnnPhys}. A detailed discussion
 on the scaling and decoupling solutions can be found in Ref.~\cite{Huber}.
In principle both, decoupling and scaling solutions are admitted by the existing lattice QCD results. Nonetheless,
 there are some indications ~\cite{Bowman,Muller} about regularity at $p^2=0$. In the present paper we
 consider the tDS equations  corresponding to the decoupling solutions.
   Recall that the rainbow approximation consists in
replacing the dressed vertices together with the dressed exchanging propagators by their bare
quantities augmented by some effective form factors

\begin{eqnarray}&&  \!\!\!\!\!\!\!\!\!\!\!\!\!\!
\left[\frac{g^2}{4\pi} \Gamma_{\mu_1 \alpha\lambda}^{(0)}(p_1,k_1,\kappa)
D^{\lambda\lambda_1}(\kappa^2)\Gamma_{\nu_1\beta\lambda_1}(p_2,k_2,\kappa)\right] =
   \Gamma_{\mu_1 \alpha\lambda}^{(0)}(p_1,k_1,\kappa) t^{\lambda\lambda_1}(\kappa)
   \Gamma_{\nu_1\beta\lambda_1}^{(0)}(p_2,k_2,\kappa) F^{eff}_{1}(p^2),
  \label{rain1}\\[3mm] &&
  \!\!\!\!\!\!\!\!\!\!\!\!\!\!
 \left [\frac{g^2}{4\pi}D_G(p^2)\Gamma_\nu(p)\right]=\Gamma_\nu^{(0)}(p)F^{eff}_{2}(p^2), \label{rain3}
 \\[3mm]  &&
 \!\!\!\!\!\!\!\!\!\!\!\!\!\!
\left[\frac{g^2}{4\pi} \Gamma_{\mu}^{(0)}(q) D^{\mu\nu}(p^2) \Gamma_{\nu}(k,q,p) \right]
= \Gamma_{\mu}^{(0)}(q)t^{\mu\nu}(p)\Gamma_{\nu}^{(0)}(k) F^{eff}_{3}(p^2),
\label{rain2}
\end{eqnarray}
where the  above three terms correspond to three  loop diagrams in Fig.~\ref{dse}.
Since we are interested  bound states, i.e. mostly in the  range of internal  momenta
 corresponding to the infra-red region, in the present paper we
 use for  $F^{eff}(p^2)$  the Gaussian form with two terms for $F^{eff}_{1,2}(p^2)$
and   one   term for $F^{eff}_{3}(p^2)$ as in~\cite{ourFB,dorkinBSmesons,OurAnalytical,EJPPlus}.
This is quite sufficient to obtain a reliable solution of the system of tDS equations.
 Such a Gaussian representation of the interaction
 kernels has been widely employed previously for quarkonia and is known as the AWW kernel~\cite{Alkofer}.
 Explicitly, in Euclidean space the effective form factors are chosen as
\begin{eqnarray} &&
F^{eff}_{1}(p^2)=  D_{11} \frac{\tilde p^2}{\omega_{11}^6} \exp{\left( -\tilde p^2/\omega_{11}^2\right)}+
D_{12} \frac{\tilde p^2}{\omega_{12}^6} \exp{\left( -\tilde p^2/\omega_{12}^2\right )} ,\label{ff1} \\ &&
F^{eff}_{2}(p^2)=  \frac{D_{2}}{\omega_2^4} \exp{\left(-\tilde p^2/\omega_2^2\right)}, \label{ff2}
\\ &&
F^{eff}_{3}(p^2)=   D_{21} \frac{\tilde p^2}{\omega_{21}^6} \exp{\left( -\tilde p^2/\omega_{21}^2\right)}+
D_{22} \frac{\tilde p^2}{\omega_{22}^6} \exp{\left( -\tilde p^2/\omega_{22}^2\right )}\label{ff11},
\end{eqnarray}
where, from now and throughout the rest of the paper, $\tilde p$ denotes the modulus of the four vector $p$ in
Euclidean space.
With such a choice of the effective interaction,  the angular integration
can be  carried out analytically~\cite{EJPPlus} leaving  h a system of
one-dimensional integral equations in Euclidean space. We found that for the set of parameters
$\omega_{11}= 1.095$ GeV, $\omega_{12}=2.15$ GeV, $D_{11}=0.465\ {\rm GeV}^2$, $D_{12}=0.116\ {\rm GeV}^2$
 for the 3-gluon loop and
$\omega_{21}= 2/3 \ \omega_{11}$ , $\omega_{22}=\omega_{12}$ , $D_{21}=0.4\pi \ {\rm GeV}^2$,
$D_{22}=0.1\pi\ {\rm GeV}^2$
for the gluon-ghost loop and $\omega_2=0.58$ GeV and $D_2=7.7 \ {\rm GeV}^2$ for the ghost loop, the solution
  the tDSE describes quite well the lattice SU(2) results.   Figures~\ref{dsress},\ref{props}  demonstrate
the obtained solution for the ghost  and  gluon   dressing functions
 and for the gluon propagator, respectively, in comparison to  the lattice
 results~\cite{BornyakovLattice,GhostLatticeMishaPRD}.

The obtained good description of the lattice data justifies  our   use of the tDSE to obtain
the gluon propagators in the complex plane
and the employment of the above effective  parameters
in solving    the tBS equation for glueballs.
\section{Rainbow approximation of the  \lowercase{t}BS amplitude} \label{bow}

By the definition, the BS pseudoscalar amplitude (\ref{BSA}) and the BS pseudoscalar vertex (\ref{vertex}) are
antisymmetric w.r.t $p_1 \leftrightarrows p_2$ and $(\mu,\nu) \leftrightarrows (\nu,\mu)$ (for a thorough analysis of   two-photon/gluon states, see Ref.~\cite{landau}).
  The most general form of such   pseudo-scalar amplitudes can be written in the form
\be
A^{\mu\nu}(p_1,p_2)=F(p_1^2,p_2^2,(p_1\cdot p_2))\epsilon^{\mu\nu\alpha\beta}p_{1 \alpha} p_{2 \beta}
\equiv F(p^2,P^2,(p\cdot P))\epsilon^{\mu\nu\alpha\beta}p_{ \alpha} P_{\beta},
\label{even}
\ee
 where the scalar functions  $ F(p^2,P^2,(p\cdot P))$, for a given glueball mass $M_{gg}^2=P^2$,
 depends solely on the relative momentum $p^2$ and the hyper angle, $\cos \xi_p$, between $p$ and $P$. Moreover, they
 are even functions of $\cos \xi_p$.  To release the tensor structure of the amplitude (\ref{even}) we multiply
 it by  $\epsilon_{\mu\nu}^{\phantom{\alpha\beta}\rho\sigma} p_\rho P_\sigma$ and contract
 all the Lorentz indices. The result is
\be &&
A^{\mu\nu}(p,P) \ \epsilon_{\mu\nu}^{\phantom{\alpha\beta}\rho\sigma} \ p_\rho P_\sigma=
2F(p^2,M_{gg}^2,\cos\xi_p)
 p_\rho P_\sigma  \left( g^{\alpha\sigma}g^{\beta\rho} -g^{\alpha\rho \beta\sigma}\right )
p_\alpha P_\beta=\nonumber \\ &&
2F(p^2,M_{gg}^2,\cos\xi_p)  \left[(p\cdot P)^2 -p^2 P^2\right]. \label{diagr}
\ee
The explicit expression for the amplitude (\ref{diagr}), within the rainbow approximation, is obtained
by  direct computation of the  (first r.h.s.) diagram in Fig.~\ref{Bseq}.
Taking into account the transversality of the amplitude, the free 3-gluon vertices
can be written as
\be   &&
\Gamma_1^{\mu_1\alpha\lambda}(p_1,k_1,\kappa)=-2g
 \left[p_1^\lambda g^{\mu_1\alpha}
-k_1^{\mu_1} g^{\alpha\lambda} -p_1^\alpha g^{\lambda\mu_1}\right],\label{G1} \\
  &&
\Gamma_2^{\nu_1\beta\lambda_1}(p_2,k_2,\kappa)=-2g
 \left[k_2^{\nu_1} g^{\beta\lambda_1} -p_2^{\lambda_1} g^{\beta\nu_1}
+p_2^\beta g^{\lambda_1\nu_1}\right]. \label{G2}
\ee
Further, we contract  the Lorentz indices in the amplitude (\ref{amplBS}) with the bare vertices
(\ref{G1})-(\ref{G2}), and the results are transformed  in to Euclidean space, where the rainbow form factors
are defined. As a result we are left with a four-dimensional integration
$d^4   k=  \tilde k^3 d\tilde k \sin^2\xi_k
d\xi_k d\Omega_{\bf k}$, where $\xi_k$ and $\Omega_{\bf k}$ are the the hyper  and spatial angles
of the momentum $k$. The scalar function  $F(\tilde p^2,M_{gg}^2,\cos\xi_p)$ in (\ref{even}) is then decomposed
over a complete set of the Gegenbauer polynomials of the first order $G_{M_p}^{(1)}(\cos\xi_p)$
 (the Chebyshev's polynomials $U_{M_p}(\cos\xi_p)$ of the second kind):
\begin{equation}
  F(\tilde p^2,M_{gg}^2,\cos\xi_p)\sin^2\xi_p=
 \sum\limits_{M_p=even} F_{M_p}(\tilde p^2,M_{gg}^2)\sin^2\xi_p G_{M_p}^{(1)}(\cos\xi_p),
\label{gegen}
\end{equation}
where the partial amplitudes $F_{M_p}(\tilde p^2,M_{gg}^2)$, for a given glueball mass
$M_{gg}$, are functions of only the relative momentum $\tilde p^2$. Calculations of the
(first r.h.s. ) diagram   in Fig.~\ref{Bseq} within the above definitions and approximations result in
\begin{eqnarray}
F_{M_p}(\tilde p^2,M_{gg}^2)&=& \frac2\pi\frac{1}{M_{gg}^2\tilde p^2}
\int\limits_0^\pi  \frac{|Z(\tilde p_1)|^2 G_{M_p}^{(1)}(\cos\xi_p)}{|\tilde p_1^2|^2} d\xi_p\nonumber \\ &\times&
\sum\limits_{M_k=even}\int \tilde k^3 d\tilde k \sin^2\xi_k d\xi_k d\Omega_{\bf k}F_{M_k}(\tilde k^2,M^2) G_{M_k}^{(1)}(\cos\xi_k)
\big  (\ldots   \big ), \label{Fk}
\end{eqnarray}
where $\tilde p_1^2=\tilde p_2^{2*} = -M_{gg}^2/4 + \tilde p^2 + i M_{gg} \tilde p \cos\xi_p
$ is the momentum of one of the constituent gluon in the Euclidean complex plane and the brackets
denote symbolically the result of contraction of the Lorentz indices in
 the expression (\ref{amplBS}). The color factor $N_c=-3$ and the corresponding powers of $(2\pi)^4$ from the
 space volume integration are also
 included into $\big  (\ldots   \big )$.
  From (\ref{amplBS}) and (\ref{even}) one infers that
the result of contracting indices is expressed in terms of some powers of four-products
$ (p\cdot P)^N=(iM_{gg}\tilde p \cos\xi_p)^N$,
$ (k\cdot P)^L=(iM_{gg}\tilde k \cos\xi_k)^L$ and $(p\cdot k)^\delta=(\tilde p\tilde k\cos\xi_{pk})^\delta$.
In total, in (\ref{Fk})
one has a five-dimensional integral. It can be essentially reduced by  observing that
the spatial dependence $\Omega_{\bf kp}$
of the integrand enters solely via the hyper angle $\cos\xi_{pk}\equiv x_{kp}=\cos\xi_p\cos\xi_k
+\sin\xi_p\sin\xi_k\cos\theta_{\bf pk}$, where $\theta_{\bf pk}$ is the spatial angle between
vectors {\bf p} and {\bf k}. There are two sources of the  hyper angle $\xi_{pk}$ dependence in (\ref{Fk}):
i) the scalar product $(p\cdot k)^\delta=(\tilde p \tilde k  x_{kp})^\delta$ which originates from
the  contractions of the Lorentz
indices and ii)    the rainbow form
factors $F^{eff}\left(\tilde p^2\right)$, which enters via the Gaussian exponents,
$\exp{\left(-(  p -   k)^2/\omega^2 \right )} =
\exp{\left(-(\tilde p^2 + \tilde k^2)/\omega^2 \right )}\cdot\exp(\alpha x_{kp}), $
where $\alpha=2\tilde k\tilde p/\omega^2$.
As a result,   the $x_{kp}$-dependence of the integrand  (\ref{Fk}) is of the
 form "$\exp(\alpha x_{kp}) x_{kp}^\delta$", which   can further be   handled  by
 decomposing it,  as above, over the same full set of the Gegenbauer  polynomials $G_{M_v}^{(1)}(x_{kp})$.

  Then the integrated over the spatial
 angles $d\Omega_{\bf k}$ of the corresponding  parts of the amplitude  can be written as
\begin{equation}
 \int  {\rm e}^{\alpha x_{kp}} x_{kp}^\delta  d\Omega_{\bf k} = 8\pi \sum_{M_v}\frac{d^\delta}{d\alpha^\delta}
 \left[\frac1\alpha I_{M_v+1}(\alpha)\right ]  G_{M_v}^{(1)}(x_k) G_{M_v}^{(1)}(x_p),  \label{integr}
 \end{equation}
where $I_{M}(\alpha)$ denotes the Bessel functions of the second kind (for details, see Appendix~\ref{A}).
 The remaining   angular integration over the hyper angle $\cos\xi_k\equiv x_k$ is of the form
\begin{equation}
{\cal K}_{M_k,M_v}^{L}  =
\int\limits_{-1}^1 G_{M_k}^{(1)}(x_k) G_{M_v}^{(1)}(x_k) x_k^L \sqrt{1-x_k^2}  dx_k,\label{intK}
\end{equation}
which can be also presented in a   closed  analytical  form (see Appendix~\ref{B}). In eq.~(\ref{intK}), $M_v$ comes from
decomposition of the rainbow exponent  (\ref{integr}),  $M_k$  from the partial decomposition (\ref{gegen})
 of the amplitude $F(\tilde k^2,M_{gg}^2, \cos\xi_k)$, and    the  term $x_k^L$ comes from
the scalar product $(k\cdot P)^L = (iM_{gg}\tilde k\cos\xi_k)^L$  which results from the
contraction of the Lorentz indices.
Then the r.h.s. of the tBS equation receives the form
\begin{eqnarray}&&
F_{M_p}(\tilde p^2,M^2)=  \frac2\pi\frac{1}{M_{gg}^2\tilde p^2}
\sum\limits_{i=1}^2\sum_{M_k=0}^{M_{max}}\sum_{M_v=0}^{M_k+3}\nonumber \\ &&
 \int\limits_0^\infty d\tilde k \tilde k^3\int\limits_{-1}^1  \frac{|Z(\tilde p_1^2)|^2 G_{M_p}^{(1)}(x_p)G_{M_v}^{(1)}(x_p)}{|\tilde p_1^2|^2\sqrt{1-x_p^2}}
 d x_p    \sum_{L,\delta}  C_{L,\delta}(\tilde k,\tilde p,x_p)
{\cal K}_{M_k,M_v}^{L} {\cal I}_{M_v}^{(\delta)}(\alpha_i) F_{M_k}(\tilde k^2,M_{gg}^2),\label{main}
\end{eqnarray}
 where
  ${\cal I}_{M_v}^{(\delta)}(\alpha_i)
 =\displaystyle\frac{d^\delta}{d\alpha^\delta}
 \left[\displaystyle\frac1\alpha I_{M_v+1}(\alpha_i)\right ] $
  with $\alpha_i=2\tilde p\tilde k/\omega_{i}^2$. Note that from a dimension analysis of the
  expression for the amplitude it follows that  $L=0...3$, $\delta=0...(3-L)$.
  The explicit expressions for the
   coefficients $C_{L,\delta}(\tilde k,\tilde p,x_p)$ can be obtained by  an analytical  manipulation
   package (e.g., Maple or Mathematica). Their explicit expressions are quite cumbersome and   are
   not presented here.
\section{Numerical calculations} \label{num}
  Expression (\ref{main}) is the main equation to be solved for the pseudo-scalar pure  glueballs.
  Recall that eq.~(\ref{main}) is written in Euclidean space, where   momenta of the constituent glueballs
  are complex and, consequently, the
  gluon propagators entering eq.~(\ref{main}) have to be defined in the complex Euclidean plane.

\subsection{Gluon propagator in the complex Euclidean plane} \label{sec:complex}
The solution of the tDS equation along the positive real axis of momenta
has to be generalized to complex values of $\tilde p_{1,2}^2$, needed
to solve the tBS equation for bound states. Note that the tDS  solutions are needed not in the the whole
Euclidean complex space, but only in the kinematical domain where
the tBS equation is defined. This is a restricted portion  of Euclidean
space which is determined  by the complex momenta of the gluon propagators $\tilde p_{1,2}^2$.  Usually this
  domain is displayed as the dependence of the imaginary part of the constituent gluon
   momentum squared, Im\,$\tilde p_{1,2}^2$,  on  its real part, Re\,$\tilde p_{1,2}^2$,
   determined by  the tBS equation.   In terms of the relative  momentum
  $\tilde p$ of the  two   gluons residing in a glueball, the corresponding dependence
  is
  \begin{eqnarray}
  \tilde p_{1,2}^2 = -\displaystyle\frac{M_{gg}^2}{4} + \tilde p^2 \pm i M_{gg} \tilde p \label{parab}
  \end{eqnarray}
   determining   in the Euclidean complex momentum plane
   a parabola~ Im\,$\tilde p^2=\pm~M_{gg}\sqrt{{\rm Re}\,\tilde p^2~+~\frac{M_{gg}^2}{4}}$ with  vertex at
  Im\,$\tilde p^2=0$ at Re\,$\tilde p^2 =-M_{gg}^2/4 $  depending  on the
  glueball mass $M_{gg}$.
In the previous  analysis~\cite{dorkinBSmesons,OurAnalytical} of the quark propagators within the rainbow
 approximation it was found  that  the  corresponding propagator functions may posses  pole-like
 singularities in the left hemisphere of the parabola,
 Re\,$\tilde p^2 < 0$,   which hamper the numerical procedure of
 solving  the tBS equation. Exactly the same situation occurs also for the gluon and ghost dressing functions
 as rainbow solutions of  tDS equations, cf. Ref.~\cite{EJPPlus}. It should be noted that
  the pole-like singularities of the propagators appear not only because of specific choice of the rainbow kernel.
 There are also some other considerations, based on studies of the gauge fixing problem, according to which
   the gluon propagator contains complex conjugate poles in the negative
 half-plane of squared complex momenta $\tilde p_{1,2}^2$, not mandatorily
 related to the rainbow approximation~\cite{ZwangerANalit,Stingl,CucchieriAnalit}.

  There are several possible procedures (cf. Refs.~\cite{analiticalFischer,GluonAnalyticalFisherPRL2012})
 of how to obtain a complex solution of the tDS equations, once the equation has
been solved for real and spacelike Euclidean momenta.
 First, one can use the so-called shell method. This method acknowledges the fact that for fixed external
momentum $\tilde p^2$ the integrand in the tDS equation samples only the mentioned parabolic domain in the complex momentum plane.
Therefore, one starts with a sample of external momenta on the boundary of a typical domain very close to
the real positive momentum axis. The tDS equations are then solved on this boundary, while the
interior points are obtained by interpolation. In the
next step, a slightly larger parabolic domain is used, with points in the interior given by the previous solution.
This way one extends the solution of the tDS equations step by step further away from the Euclidean
result into the whole complex plane. A shortcoming of the method is that there is an accumulation
 of numerical errors at each step of the calculations.

A second option  is to deform the loop integration path itself away from the real positive
$\tilde p_{1,2}^2$ axis~\cite{OurAnalytical,MarisComplexDSE}.  This can be done by deforming the integration
contour and solving the integral equation along this new contour.
In practice, one changes the integration contour by rotating it in the complex plane,
multiplying both the internal and the external variables by a phase factor $e^{i\phi}$. Thus,    one gets the
complex variables $\tilde p=|\tilde p| e^{i\phi}$  and  $\tilde k =|\tilde k| e^{i\phi}$
 and solves the tDS equation along the rays $\phi=const$.
This method works quite well in the first quadrant, $\phi \le \pi/2$,
but fails at $\phi > \pi/2$, see e.g. Refs.~\cite{OurAnalytical,dorkinBSmesons}.
 This is because along the rays $\phi=const$
all the values of $|\tilde p|$, from $|\tilde p|=0$  to $|\tilde p|\to\infty$ contribute to the tDS equation,
  even if  one needs the solution only in a  restricted area of the parabola
 Re\,$\tilde p_{1,2}^2 < 0$. Consequently, numerical instabilities are inevitable at $\phi > \pi/2$.

The third method, which we use in this work, consists in  finding a solution
of the integral equations in a straightforward way from the tDS equation along the real $\tilde k_{1,2}$ axis
  on a complex grid for the external momentum $\tilde p_{1,2}$ inside and on the
  parabola (\ref{parab}).
  As in the previous case, numerical instabilities  in the tDS equation can be caused by oscillations of
  the exponent ${\rm e}^{-(\tilde k_{1,2}-\tilde p_{1,2})^2/\omega^2}$
  at large $|\tilde p_{1,2}^2|$ and/or at large  $\tilde k_{1,2}^2$.  However, one can get rid of such
   a numerical problem by taking into account
  that the parabola (\ref{parab}) restricts only a small portion of the complex plane at
  Re\,$p_{1,2}^2<0$, where  the numerical problems are minimized.
 For positive values of Re\,$\tilde p_{1,2}^2>0$, where $|\tilde p_{1,2}^2|$ can be large, i.e.
the  relative momentum $\tilde p$ in (\ref{parab}) can be  large,
  the tBS wave function of a glueball  is expected to decrease rapidly with increasing values  of its
  argument $ \tilde p$, and
  at $\tilde p^{max}\sim 3-4 \ {\rm GeV/c}$  to become  already sufficiently small. In such manner, one can
 solve the complex  tDS equation at not too large values of  $|\tilde p_{1,2}^2|$, where  a reliable calculation
 of the loop integrals   is still possible. Then one takes advantage
of the fact that, at larger values of $\tilde p$, the highly oscillating integrals,
 in accordance with the Riemann-Lebesgue lemma,
are negligibly small or  even vanish at $\tilde p \to\infty$.
Consequently, one can either completely neglect the contribution to the propagators in this region
or   use a simple asymptotic parametrization of the real propagators and continue it in the complex plane.
In the present paper we use the latter option with explicit  parametrizations of lattice
 data~\cite{BornyakovLattice}
to which our effective parameters have been adjusted.
Note that   attempts to extend  the parametrizations of the lattice data from the positive
 momenta to the left hemisphere of   Euclidean plane are inconsistent, since such a procedure
 can lead to essentially different results
 differing by orders of magnitudes from each other, see e.g.~\cite{param}.
 However, for large  positive  momenta, such an analytical continuation is applicable. In the present paper
 the complex gluon propagators are found by solving the tDS equations in the
 the left hemisphere Re $\tilde p_{1,2}^2 <0$ and in a part of the right hemisphere
 Re $\tilde p_{1,2}>0$ determined by the integration momentum $0\le  \tilde k \le (6-7)$~GeV;
 for the remaining parabolic domain we use the explicit parametrisation
 of lattice data from Ref.~\cite{BornyakovLattice}.

\subsection{Ingredients for the determinant} \label{numBS}

Having fixed the complex gluon propagators, the  integration over the momentum $\tilde k$ is executed by
discretizing the integral by a proper  choice of the Gaussian mesh. The integration interval
 $\tilde k=[0,\infty]$  is truncated by a sufficiently large value of $\tilde k ={\cal O}(25)$~GeV.
Within this  interval, the gluon propagators are determined by solving the tDS equations
for $0\le\tilde k < \left( \displaystyle\frac{M_{gg}}{2} + 6\right)$~GeV, and by using the parametrization
of lattice data~\cite{BornyakovLattice} for larger values of $\tilde k$.
In such a way, the tBS equation for the  amplitude transforms in to a homogeneous system of algebraic equations of the
form

\begin{equation}
X= \, S X, \label{syst}
\end{equation}
where  the vector
\begin{equation}
X^T=\left[ F_{Mp=0}(\tilde p_1), F_{Mp=0}(\tilde p_2),\ldots F_{Mp=0}(\tilde p_{N_G}),\ldots
F_{Mp=M^{max}}(\tilde p_1), F_{Mp=M^{max}}(\tilde p_2),\ldots F_{Mp=M^{max}}(\tilde p_{N_G})
\right ],
\label{vector}
\end{equation}
\noindent
 for a given   value of $M_{gg} $,
represents the sought solution in the form of a group of sets of
partial wave components $F_{M_p}(\tilde p_i)$,
  specified on the integration mesh of the order $N_G$ and the maximum number $M^{max}$ of the Gegenbauer
 polynomials used in (\ref{gegen}). In our calculations we use $M^{max}=4-5$, i.e. the Gegenbauer
 polynomials, which must be even functions of their arguments, run from $G_{M_p=0}^{(1)} (x_p)$
 to $G_{M_p=6,8}^{(1)}(x_p)$. Actually, we found that already for $M_p=6$ the convergence of the solution
 is rather good. However, the final results are obtained for $M^{max}=5$, i.e. the maximum order of
 the Gegemnbauer polynomial in (\ref{gegen}) is $M_p=8$.
   The resulting matrix $S$ is of   dimension $N_S\times N_S$, where $N_S=N_G \times  M ^{max}  $. In our
   calculations we use a Gaussian mesh with $64$ nodes in the left hemisphere of the parabola and $96$
   for the rest of the integration domain. In total the Gaussian mesh in our calculations consists  of $160$ nodes, so
   that the dimension of the matrix $S$ is $800\times 800$ which is not too large to obtain reliable numerical
   results.
Since the system  (\ref{syst}) is homogeneous,
the  eigenvalue  solution  is obtained from the condition $\Delta=\det(S-\mathbb I)=0$.
More details about the numerical algorithms of solving the BS equation can be found elsewhere, cf.
Refs.~\cite{ourCiofi,dorkinBSmesons,dorkinBeyer}.

 \section{Results}
\begin{figure}[!ht]
  \begin{center}
 \includegraphics[scale=0.4 ,angle=0]{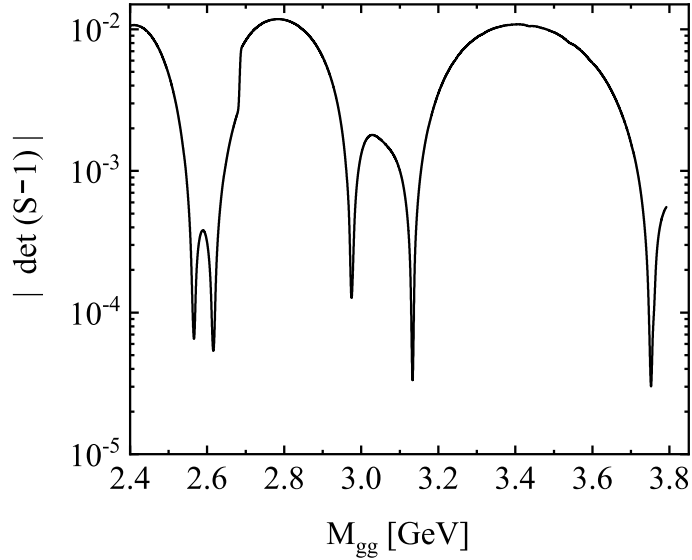}
  \end{center}
 \caption{ The  absolute value of the determinant of the matrix  $ (S-\mathbb I)$ in dependence of
  the glueball mass $M_{gg}$. The deep minima correspond to pure glueball bound states.
  Ideally, the minima should be proper zeros,
however, the finite numerical resolution prevents this. Since the determinant changes its sign at
the minima, we attribute them to zeros.}
 \label{deter}
 \end{figure}

 The solutions of the tBS equation (\ref{main}) or, equivalently the solutions of
 eq.~(\ref{syst}), are sought as   zeros of the determinant  of the matrix $ (S-\mathbb I)$.
We scan the values of  $M_{gg}$ from a minimum value $M_{gg}\sim 0.1$~GeV to a
maximum value $M_{gg}\sim 4$~GeV with a scanning step of $8-10$~MeV.
At each stage we compute the corresponding  determinant  and look for the change of the sign,
which clearly  would  indicate   that in the neighbourhood of this interval the determinant has a zero,
 i.e. this is the sought interval where  the solution of tBSE is located.
The matrix elements of $S$ are computed with the same set of
effective parameters as used in solving the tDS equations for gluon and ghost propagators and which
assures a good description of the lattice data, cf. Figs.~\ref{dsress}~and~\ref{props}. In the decomposition
of the amplitude over the Gegenbauer polynomials (\ref{gegen}) we take into account up to five terms, i.e.
$M_p=0,2,4,6$~and~$8$. The method converges already for 3-4 terms in (\ref{gegen}),
however for a more stable results we included   also $M_p=8$. We found that the first
zero of the determinant, i.e. the solution for the pure glueball ground states,
corresponds to $M_{gg}=2560$~MeV
which is quite close to the predictions by the lattice calculations
$M_{gg}^{(0)}=2590\pm136$~MeV~\cite{Chen,Morningstar}.
Next three zeros have been  found to be located at
$M_{gg}^{(1)}=2620, M_{gg}^{(2)} =2973$ and $M_{gg}^{(3)}=3130$~MeV,
which do not have an analogue
with lattice data. The next zero at $M_{gg}^{(4)}=3745$~MeV is quite close to
 the first  excited state
predicted by lattice calculations, $M_{gg}^{({\rm 1st.})}=3640\pm 189$~MeV.
As an illustration, in Fig.~\ref{deter} we present the behaviour of the absolute value
of the determinant as a function of the
mass $M_{gg}$ of two dressed gluons in the interval $M_{gg}=2.4 - 3.8 $~MeV where
the zeros of the determinant have been detected, i.e. where the   bound states occured.

Thus, we see that
the obtained mass spectrum is more rich than the one predicted by
lattice calculations~\footnote{It should be noted that the most recent publication~\cite{Chen} does not longer consider
the excited states, so that the state $M_{gg}^{({\rm 1st.})}=3640\pm 189$~MeV should be considered
with some caution. }.
This is not a  new results in investigations of  the energy/mass spectra within relativistic equations.
Usually, the corresponding equations provide much more states than
the observed real experimental spectrum. Some solutions are in a sense redundant.
It is well known that,  for quark-antiquark bound
states, the  tBS equations posses solutions for some combinations $q\bar q$ quarkonia which
do not exist in nature. For instance, for the pseudo-scalar  $q\bar q$ states,
the tBS approach exhibits solutions for $s\bar s$, $c\bar s$ etc.~\cite{dorkinBSmesons} which are not
detected experimentally. On the other side, the mass spectrum of identified mesons is reproduced
within   tBS approach with a  very good accuracy~\cite{HilgerQuarkonia}, i.e. the real meson spectrum is
entirely contained in the spectrum of the numerical solutions of  tBS . Yet, nowadays in the  literature one
starts to discuss the so-called "abnormal" BS solutions, firstly reported as Wick-Cutkosky
amplitudes for the BS equation with interaction kernel mediated by
 exchange of massless particles~\cite{karmanov}. It is  also found that in case of massive
exchanging particles  some solutions of the  (very simplified) BS equations
 disappear in the non-relativistic limit for
  the speed of light $c\to \infty$, i.e., presumably such abnormal states cannot be observed
experimentally~\cite{karmanov}.

It should be noted furthermore  that the lowest lying glueball
states have been considered, in a consistent  manner,  in Ref.~\cite{glueBS} where
 a Dyson-Schwinger-Bethe-Salpeter approach has been employed. As in the present paper, the   gluon
 propagators entering the tBS scheme have been
 taken  as solutions of the previously solved tDS approach~\cite{PawlowskyFicher,GluonAnalyticalFisherPRL2012}
 and  generalized to the complex Euclidean plane. Particular
 attention in ~\cite{PawlowskyFicher,GluonAnalyticalFisherPRL2012} was
 paid to parametrizations of  the three-gluon vertex and the ghost-gluon
 vertex to satisfy the Slavnov-Taylor identity. The obtained results show a good agreement of the
 calculated scalar ($0^{++}$) glueball mass ($M_{gg}=1.64\ {\rm GeV}$)
 with the lattice data ($M_{lat.}=(1.73\pm 0.094)\ {\rm GeV}$), while the computed mass of
  the pseudo-scalar ($0^{-+}$)
 glueball ($M_{gg}=4.53\ {\rm GeV}$) is almost twice larger than the lattice
 predictions ($M_{lat.}=(2.590\pm 0.136)\ {\rm GeV}$). Also, the pseudo-scalar glueballs have been
recently considered within the rainbow approximation of the tBS equation in Ref.~\cite{roberts20New}, where the dependence of the
 glueball mass on the effective slope parameters has been investigated. However, the relevance of the effective
rainbow  parameters to the gluon and ghost propagators, as well as to the lattice results, has not been
discussed.

\begin{figure}[!ht]
  \begin{center}
 \includegraphics[scale=0.3 ,angle=0]{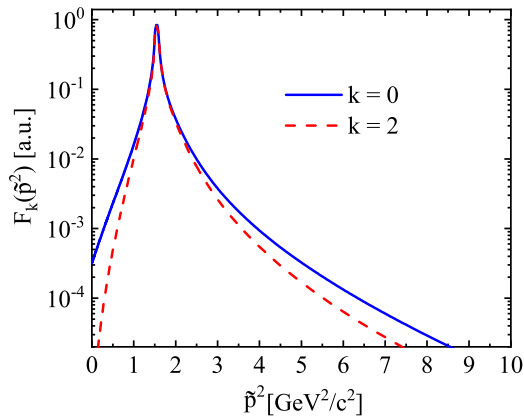}
  \end{center}
 \caption{ The first two partial amplitudes (Chebyshev's modes), eq.~(\ref{gegen}), as functions of the
 Euclidean momentum $\tilde p^2$. The presented amplitudes correspond to the BS amplitude for
 the pseudo-scalar glueball ground  state $M_{gg}=2560$~MeV.}
 \label{partial}
 \end{figure}

Having found the values of masses of dressed pure two-gluon states  which assure the
compatibility of the system (\ref{syst}),
one can straight forwardly find the partial BS amplitudes (\ref{vector}). Since the tBS equation
 is a homogeneous equation,
the amplitude can be found up to an arbitrary constant. We solved the system (\ref{syst}) and normalized the
partial amplitudes to the maximum value of the first term, $F_0(\tilde p^2,M_{gg}^2) $,
which occurs at $\tilde p^2 \sim 1.57\ {\rm GeV}^2 $. It turns out that in the region of the maximum
the first two amplitudes,
$F_0(\tilde p^2,M_{gg}^2)$ and $F_2(\tilde p^2,M_{gg}^2)$, are basically of the same magnitude,
while the subsequent terms $F_4(\tilde p^2,M_{gg}^2)$, $F_6(\tilde p^2,M_{gg}^2)$ and
$F_8(\tilde p^2,M_{gg}^2)$ are essentially smaller, each being smaller than its previous
neighbour  by a factor $\sim 3$.
For instance, at $\tilde p^2\simeq 6.5\ {\rm GeV}^2 $ the amplitudes
$F_8(\tilde p^2,M_{gg}^2)$ is by more than   two order of magnitudes smaller than $F_0(\tilde p^2,M_{gg}^2)$.
 Also, the amplitudes $F_0(\tilde p^2,M_{gg}^2)$ and $F_4(\tilde p^2,M_{gg}^2)$
are positive in the whole kinematical range, the amplitude $F_2(\tilde p^2,M_{gg}^2)$ is negative everywhere,
while the other amplitudes are not of a definite sign. As an illustration of the behaviour of the partial
amplitudes,   in Fig.~\ref{partial} we present   the main two amplitudes, $F_0(\tilde p^2,M_{gg}^2)$ and
$\left |F_2(\tilde p^2,M_{gg}^2)\right |$, as functions of the Euclidean relative
momentum $\tilde p^2$. It is seen that both  amplitudes are mainly located around their maximum
at $\tilde p^2  \simeq 1.57\ {\rm GeV}^2$ sharply decreasing  away from the maximum location. Such a
$\delta$-function like behaviour is observed for the remaining
$F_4(\tilde p^2,M_{gg}^2)$ - $F_8(\tilde p^2,M_{gg}^2)$ partial amplitudes as well. The behavior of the
partial amplitudes of the exited stats are basically identical to the ones for the ground state, except that
 the maximum
is shifted towards larger $\tilde p^2$. Similar qualitative behaviour of the zero's Chebyshev mode has
been recently reported in Ref.~\cite{roberts20New}.
\section{Summary} \label{summary}
In summary, we present a rainbow approximation to the Dyson-Schwinger-Bethe-Salpeter approach to
analyse the spectrum mass of   pseudo-scalar glueballs. We argue that it is possible to determine a set
of effective parameters which describes fairly well the gluon and ghost propagators from the
truncated Dyson-Schwinger equation in comparison to the lattice results. The same set of parameters provides
the solutions of the Bethe-Salpeter equation for the mass spectrum of the pseudo-scalar glueballs. It is shown that
the obtained mass spectrum includes the ground and first excited states predicted by lattice calculations. Besides,
in the interval $M_{gg}=2.5 - 4$ GeV there are more states then predicted by lattice calculations.
This is usual situation when solving the equations, non-relativistically and relativistically,
for the bound state energies. Some states could be redundant,
other ones can belong to the so-called "relativistic abnormal" states,
which disappear in the non relativistic limit, i.e. cannot be detected experimentally.  However, the theoretical
description of these abnormal states is not yet firmly  settled and we do not discuss it in details here.

\section*{Acknowledgments}
This work was supported in part by the Heisenberg - Landau program
of the JINR - FRG collaboration, GSI-FE and BMBF. LPK appreciates the warm hospitality at the
Helmholtz Centre Dresden-Rossendorf.

 \appendix
 \section{Partial Decomposition of the rainbow kernel} \label{A}
 The spatial dependence of the integrand on $\Omega_{\bf k}$ is contained in the rainbow  exponents and in the
 scalar product $(p\cdot k)^\delta= (\tilde p\ \tilde k \ x_{kp})^\delta$,

 \begin{eqnarray}&&
 \exp(\alpha x_{kp}) x_{kp}^\delta = \sum_{M_v} W_{M_v}^{(\delta)}(\tilde p,\tilde k) G_{M_v}^{(1)}(x_{kp}),
 \label{pot}
 \end{eqnarray}

\noindent
 where $\alpha=2\tilde k\tilde p/\omega^2$. The partial coefficients $W_{M_v}^{(\delta)}(\tilde p,\tilde k) $
 can be computed explicitly as
 \begin{eqnarray}&&
W_{M_v}^{(\delta)}(\tilde p,\tilde k)=\frac2\pi\int_{-1}^1 \sqrt{1-x_{kp}^2} \exp(\alpha x_{kp}) x_{kp}^\delta
 G_{M_v}^{(1)}(x_{kp})d x_{kp}=\nonumber \\ &&
 \frac2\pi\frac{d^\delta}{d\alpha^\delta}\left[ \int_{-1}^1 \sqrt{1-x_{kp}^2} {\rm e}^{\alpha x_{kp}}
 G_{M_v}^{(1)}(x_{kp})\right] d x_{kp}
  = 2(M_v+1)    \frac{d^\delta}{d\alpha^\delta}
  \left[\frac1\alpha I_{M_v+1}(\alpha)\right ],
 \end{eqnarray}
where $I_{M_v+1}(\alpha)$ are the modified Bessel functions of second kind (of the imaginary argument) yielding

 \begin{eqnarray}
{\rm e}^{\alpha x_{kp}} x_{kp}^\delta =2 \sum_{M_v} (M_v+1)
\frac{d^\delta}{d\alpha^\delta}\left[\frac1\alpha I_{M_v+1}(\alpha)\right ]
   G_{M_v}^{(1)}(x_{kp}).
 \end{eqnarray}
The dependence on the spatial angles of the vectors $\bf p$ and ${\bf k}$  enters via $G_{M_v}^{(1)}(x_{kp}\equiv\cos\xi_{kp})$,
 where $\cos\xi_{kp}=\cos\xi_p\cos\xi_k + \sin\xi_p\sin\xi_k\cos\theta_{\bf kp}$. Explicitly, such a dependence can be written
 by using an addition theorem for Gegenbauer polynomials
 \begin{eqnarray}
  G_{M_v}^{(1)}(x)=\frac{2\pi^2}{M_v+1} \sum_{l\mu} Z_{M_v l\mu}^*(p) Z_{M_v l\mu}(k)
  \end{eqnarray}
with $ Z_{M_v l\mu}(k)=Z_{M_v l\mu}(\xi_k,\theta_{\bf k},\phi_{\bf k})$ as hyper-spherical harmonics, to  obtain
\begin{eqnarray}
{\rm e}^{\alpha x_{kp}} x_{kp}^\delta =4\pi^2 \sum_{M_v,l,\mu} \frac{d^\delta}{d\alpha^\delta}
\left[\frac1\alpha I_{M_v+1}(\alpha)\right ] Z_{M_v l\mu}^*(p) Z_{M_v l\mu}(k),\label{dvas}
\end{eqnarray}
where the normalized hyperspherical harmonics are $Z_{M_v l\mu}(p) = X_{M_v l}(\xi_p) {\rm Y}_{l\mu}({\bf p})$ with
\begin{equation}
X_{M_v l}(\xi_p) = 2^l l!\sqrt{\frac2\pi} \sqrt{\frac{(M_v+1)(M_v-l)!}{(M_v+l+1)!}}
\sin^l\xi_p G_{M_v-l}^{l+1}(\cos\xi_p). \label{dvass}
\end{equation}
At a first glance, equations (\ref{pot})-(\ref{dvass}) seemingly even complicate  the integration. However,
 by observing that the dependence
of the integrand  in (\ref{amplBS}) on the spatial angles $\Omega_{\bf k}$ is only through the interaction
kernel and trough $x_{kp}^\delta$, eq. (\ref{dvas}), i.e. only trough the spacial
 harmonics ${\rm Y}_{l\mu}({\bf k})$, the integration over
$d\Omega_{\bf k}$ is trivial and eventually we have
\begin{equation}
 \int  {\rm e}^{\alpha x_{kp}} x_{kp}^\delta  d\Omega_{\bf k} = 8\pi \sum_{M_v}\frac{d^\delta}{d\alpha^\delta}
 \left[\frac1\alpha I_{M_v+1}(\alpha)\right ]  G_{M_v}^{(1)}(x_k) G_{M_v}^{(1)}(x_p).  \label{integr1}
 \end{equation}
 \section{Integrations over $x_k$} \label{B}

  Hereinbelow we present some details of integrations over the hyper angle $x_k$ and the
  resulting  explicit expressions of selection rules.
The corresponding angular  integral is of the form
   \begin{eqnarray}
   {\cal K}_{M_v,M_k}^{L}=   \int\limits_{-1}^1\sqrt{1-x_k^2} x_k^L G_{M_k}^{(1)}(x)G_{M_v}^{(1)}(x_k) dx_k.
   \label{integr2}
   \end{eqnarray}
   Due to parity
   restrictions, the  partial amplitudes $F_{M_k}$ contain only even   values
   of the Gegenbauer polynomials, i.e. $  \ M_k=[0,2,4,.. M^{max}]$,
   where $ M^{max}$ is the maximum number of polynomials taken into account in concrete calculations.
    The Gegenbauers $G_{M_v}^{(1)}(x_k)$ which come from the interaction kernel (\ref{integr1})
  may   contain both, even and odd values of $M_v$, and formally the summation
  is extended  to infinite, $M_v=[0..\infty]$.
     However, not all values in this interval contribute to (\ref{integr2}). The symmetrical
      limits of integration
      restrict the Gegenbauer polynomials in (\ref{integr2}) to obey    the condition
   ($L+M_k+M_v$)=even. Other restrictions originate
   from the explicit expression of the integral, see below.
   From a standard math handbook one infers that

 \begin{eqnarray} &&
  \int\limits_{-1}^1\sqrt{1-x^2} x^L G_{M_k}^{(\lambda)}(x)G_{M_v}^{(\lambda)}(x) dx =
  2^{M_k+M_v}\frac{(2\lambda)_{M_k} (2\lambda)_{M_v} L!}{m_k!M_v!(L+M_k+M_v)!}\left(\frac{L-M_k-M_v}{2}+1\right)_{M_k+M_v}
  \nonumber\\ &&
  \times B\left(\lambda+\frac12,\frac{L+M_k+M_v+1}{2}\right)
  \!\! \phantom{1}_3F_2\left(-M_k,-M_v,1;2\lambda,\frac{L-M_k-M_v}{2}+1;1\right),
  \label{prud}
  \end{eqnarray}
where  $ \!\! \phantom{1}_3F_2$ is the generalized hypergeometric function
and $(a)_k=a (a+1)(a+2)\ldots(a+k-1)$ (with $a_0=1$) is the known Pochhammer symbol and
 $B(x,y)=\displaystyle\frac{\Gamma(x)\Gamma(y)}{\Gamma(x+1)}$ and $\Gamma(x)$ are
  the familiar Euler $\beta-$ and $\Gamma-$ functions, respectively.
Despite the integral  (\ref{prud}) is finite,  at some  values of  $L,M_k$ and $M_v$ the
product of the Pochhammer symbol and hypergeometric function
can be of the type $0 \cdot \infty$, which implies that Eq. (\ref{prud})
cannot be implemented directly in to   numerical calculations.
One needs to handle zeros and singularities manually. We use the  obvious properties
\begin{eqnarray}
 (-m)_k=(-1)^k\frac{ m!}{(m-k)!}; \qquad
(a)_k=\frac{\Gamma(a+k)}{\Gamma(a)}  \label{obv}
\end{eqnarray}
  to obtain
  \begin{eqnarray} &&
\left(\frac{L-M_k-M_v}{2}+1\right)_{M_k+M_v}=\frac{\kappa!}{\kappa_1!};\\[1mm] &&
B\left(\frac32,\frac{L+M_k+M_v+1}{2}\right)=\frac{\pi}{2^{\kappa+1}}\frac{(2\kappa)!!}{(\kappa+1)!}; \\[1mm] &&
\!\! \phantom{1}_3F_2\left(-M_k,-M_v,1;2,\frac{L-M_k-M_v}{2}+1;1\right)=
\sum_{k=0}^\infty \frac{1}{(k+1)!}  \frac{M_k!M_v!}{(M_k-k)!(M_v-k)!}
\frac{\kappa_1!}{(\kappa_1+k)!},\nonumber \\
  \end{eqnarray}
  where, for   brevity, we introduce the shorthand notation $\kappa=\displaystyle\frac{L+M_k+M_v}{2}$,
 $\kappa_1=\displaystyle\frac{L-M_k-M_v}{2}$.

With this notation, the integral (\ref{prud}) reads
\begin{eqnarray} &&
 \int\limits_{-1}^1\sqrt{1-x^2} x^L G_{M_k}^{(1)}(x)G_{M_v}^{(1)}(x) dx =\pi
 \frac{2^{M_k+M_v-\kappa-1} (M_k+1)!(M_v+1)!L!}{(\kappa+1)(2\kappa)!!}\times \nonumber \\&&
 \sum_{k=0}^\infty \frac{1}{(k+1)!(M_k-k)!(M_v-k)!(\kappa_1+k)!}.
 \label{resul}
 \end{eqnarray}
   In  eq.~(\ref{resul}) the   summation is  restricted by those values of $k$ which
   ensure   non-negative factorials,
 i.e. in the above sum $k\le M_k$ and $k\le M_v$ and $(M_k+M_v-L+2k)\ge 0$.
 Together with the condition $(L+M_k+M_v)$-even, these
 restrictions form the selecting rules for the  integral (\ref{integr2}). Actually,
 in practice the summation in (\ref{resul}) consists only of one, or maximum two terms. Consequently,
the integrals (\ref{integr2})   turn  out to be
 extremely  simple being expressed in form of the fractional parts of $\pi$.
 For instance, the value  {  L=0}   results in the orthogonal condition for the Gegenbauer polynomials
 i.e.   ${\cal K}_{M_k,M_v}^{L=0}=\displaystyle\frac\pi2\delta_{M_k,M_v}$. For $L=1$ one has
 $M_v=1,3,5,7$ and  for even $M_k=0,2,4,6,8$ the integral (\ref{integr2}) is always $\pi/4$, etc.

 \end{document}